# Con los juegos también se educa: un enfoque educativo de los juegos de la Oca y el Parchís

Luis Alvarez León[1], Pablo García Tahoces[2] y Emilio Macías Conde[1]

[1]CTIM. Departamento de Informática y Sistemas, Universidad de Las Palmas de Gran Canaria, Campus de Tafira, 35017 Las Palmas de G.C., Spain

[2]CITIUS. Centro de Investigación en Tecnoloxías da Información, Universidad de Santiago de Compóstela, Campus Vida, 15782 Santiago de Compostela, Spain

Email : lalvarez@ctim.es, pablo.tahoces@usc.es

## RESUMEN

De acuerdo con el informe Horizon Internacional del 2011 (tradicional diagnóstico y pronóstico del uso de tecnologías y tendencias educativas de futuro), el aprendizaje basado en juegos será una de las áreas de mayor crecimiento en los próximos años en el contexto de la aplicación de las nuevas tecnologías a la enseñanza. En el año 2003, James Gee desarrolló algunos trabajos sobre el impacto del juego en el desarrollo cognitivo. Desde entonces, la investigación y el interés en el potencial de los juegos en el aprendizaje se ha disparado, dando lugar a una nueva área de trabajo denominada "serious games", donde el juego es un vehículo para mejorar y facilitar el aprendizaje. Los niños nacidos a partir de 1990, han crecido en un mundo donde han tenido acceso a todo tipo de dispositivos electrónicos (ordenadores, tabletas, móviles, etc..). Explotar este conocimiento "digital" de los niños para incorporarlo a los procesos de aprendizaje es algo de gran interés educativo.

En este trabajo se muestra como, a partir de juegos de sobremesa tradicionales como el juego de la Oca o el Parchís, es posible diseñar juegos educativos que tengan como objetivo reforzar el aprendizaje de los niños. La idea que vamos a desarrollar para hacer esto es muy simple : los niños juegan con las mismas reglas que en el juego tradicional, añadiendo además la funcionalidad de que después de tirar el dado se formula una pregunta, de tal manera que la ficha solo se mueve en el caso en que la pregunta se responda correctamente. Las ventajas que tiene este enfoque son las siguientes :

1. La gran mayoría de los niños ya está familiarizado con las reglas de estos juegos.

2. Se fomentan los juegos/trabajos en equipo.

3. Modificando la base de datos de preguntas, el aprendizaje se puede adaptar a muchas situaciones distintas.

En la página web : http://www.ctim.es/SeriousGames/ puede encontrarse una implementación práctica de estas ideas.

# INTRODUCCIÓN

La mayoría de nosotros está familiarizado con los juegos de mesa clásicos como el Parchís o el Juego de la Oca. Así que, ¿por qué no utilizar estos juegos como base para crear juegos educativos, añadiendo algunas funcionalidades extra? El objetivo principal de este trabajo es explorar cómo agregar funcionalidades educativas a los juegos de mesa clásicos, pero manteniendo el "espíritu" original del juego.

Hemos estudiado en detalle algunos juegos de mesa clásicos como el Juego de la Oca, el Parchís y también hemos diseñado uno nuevo: "El Juego del Motor" con una mecánica similar pero inspirado en las carreras de coches (por ser éste un tema atractivo para los niños). Todos estos juegos tienen las siguientes características comunes:

1. En el participan varios jugadores (o eventualmente equipos).

2. Cada equipo tiene uno o varias fichas que se mueven en un tablero lanzando un dado.

3. La forma en que una ficha se mueve en el tablero sigue reglas simples.

4. Cada ficha se inicia en una posición determinada y sigue una ruta única para llegar a una posición final (los jugadores no pueden elegir dónde mover el ficha).

5. El equipo ganador es el que llega primero a la posición final.

Teniendo en cuenta estas características es fácil diseñar una aplicación informática para realizar automáticamente el movimiento de la ficha de acuerdo con las reglas de juego y el lanzamiento del dado. Pero queremos ir más allá en este análisis. Queremos añadir funcionalidades educativas a los juegos. Esto lo haremos de una forma muy simple: al comienzo del juego, los jugadores eligen uno o varios temas y durante el juego, para poder mover sus fichas, los jugadores tienen que responder a preguntas sobre estos. Es decir, antes de mover la ficha, el jugador (o su equipo) tiene que responder correctamente a una pregunta elegida al azar por el sistema en la base de datos de preguntas. Los temas y las preguntas de cada tema pueden ser fácilmente modificados para adaptar los juegos a contextos de aprendizaje muy diferentes. También se ha incluido la opción de añadir imágenes a las preguntas, lo que aumenta mucho su potencial educativo.

La organización de este trabajo es la siguiente: En la sección 2 se estudia el diseño del juego educativo. En la sección 3 se muestran algunos resultados obtenidos y, finalmente, en la sección 4 se presentan las principales conclusiones del trabajo.

# DISEÑO DE LOS JUEGOS EDUCATIVOS

Las motivaciones/requisitos que teníamos en mente cuando estudiamos la forma en que se podría transformar los juegos de mesa clásicos en juegos educativos fueron las siguientes:

1. Aprovechar que muchas personas están familiarizadas con los juegos de mesa clásicos tomándolos como base para hacer juegos educativos atractivos y fáciles de usar.

2. Mantener el espíritu de juego de mesa original en términos de las reglas básicas del juego.

3. El nuevo juego debe ser fácilmente adaptable a diferentes contextos de aprendizaje para poder ser utilizado en diferentes niveles educativos y con diferentes objetivos.

4. El nuevo juego debe funcionar en el mayor número posible de sistemas informáticos (Ordenadores Windows - Mac - Linux, tabletas, etc..)

5. El nuevo juego debe poder ejecutarse tanto a través de un servidor en internet como localmente en un ordenador sin conexión a internet.

6. El nuevo juego debe añadir la opción de acelerar el tiempo de juego y hacer partidas más rápidas cuando se dispone de poco tiempo.

7. El software informático debe poder ser modificado por terceros para adaptarlo a sus necesidades.

### Selección de juegos

Hemos elegido 2 juegos de mesa clásicos: Juego de la Oca y Parchís, y hemos diseñado uno nuevo "El Juego del Motor", con una mecánica similar, pero ambientado en carreras de coches. A continuación se hace un breve resumen de las características de los juegos:

1. **Juego de la Oca**: cada jugador/equipo tiene una sola ficha. Las fichas se mueven por el tablero tirando un dado. Para ganar los jugadores tienen que llegar a la casilla de la meta con un número exacto al tirar los dados (es decir si se saca un número mayor al necesario la ficha retrocede).

2. **Parchís**: Técnicamente es más complejo que la Oca. Los jugadores/equipos juegan con cuatro fichas y ellos pueden decidir, en cualquier parte, la ficha que se mueve. Hay un número de posibles interacciones entre las fichas en el tablero que se tienen que tomar en cuenta. Los equipos tienen que llegar a la meta con un número exacto al tirar los dados.

3. **El Juego del Motor**: es muy sencillo, con una mecánica similar a la Oca. Cada jugador/equipo tiene una sola ficha, las reglas son simples. Para ganar no es necesario alcanzar la meta con un número exacto al tirar los dados (basta con que el número obtenido sea superior al necesario para llegar a la meta).

**Adición de funcionalidades para añadir una función educativa a los juegos.**

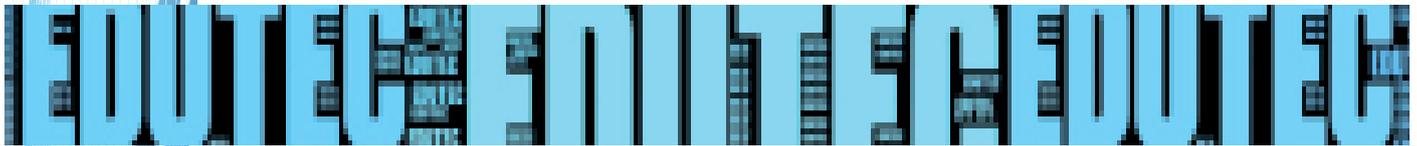

La forma en que se decidió incluir funcionalidades educativas a los juegos fue agregar preguntas a las que los jugadores/equipos tienen que responder antes de mover sus fichas. Al comienzo del juego, se eligen los temas de las preguntas que se van a realizar. A cada equipo se le asignan uno o varios temas que además pueden ser diferentes para los equipos. Dado que los temas pueden ser variados para los diferentes equipos, es posible jugar con grupos de niños de diferentes edades, adaptando la dificultad de los temas a su edad. Por ejemplo, en la base de datos de preguntas en inglés utilizada en esta primera versión de la aplicación, se han incluido los siguientes temas:

1. Comida: identificar los alimentos por sus imágenes asociadas
2. Animales (nivel infantil): identificar las mascotas o animales muy conocidos a través de imágenes.
3. Animales: Lo mismo que el anterior pero con animales menos conocidos.
4. Deporte: Identificar los deportes por imágenes de sus símbolos
5. Señales de la vida cotidiana: Identificar el significado de las diferentes señales que podemos ver en la calle.

Durante el juego, para poder mover sus fichas, los jugadores tienen que responder a preguntas sobre estos temas elegidos previamente al comenzar la partida.

### Gestión de la base de datos de preguntas.

Un tema importante que se ha abordado en este trabajo es estudiar un procedimiento para que un usuario, sin conocimientos de programación, (por ejemplo un profesor) pueda cambiar/añadir preguntas y temas a la base de datos. La estructura interna de la base de datos se gestiona utilizando javaScript (White, 2009). Para evitar que el usuario tenga que manejar directamente código de javaScript, se ha ideado un procedimiento donde a partir de una hoja de cálculo con los datos de las preguntas, se generan automáticamente los ficheros javaScript con las preguntas. De esta manera los pasos para agregar / modificar los temas de las preguntas son los siguientes:

1. Modificar la hoja de cálculo que se suministra como ejemplo para añadir nuestros propios temas y preguntas.

2. Ejecutar un programa que se suministra para convertir la hoja de cálculo en los ficheros javaScript asociados.

3. Reemplazar los archivos javaScript existentes con los nuevos en el directorio adecuado (teniendo en cuenta el idioma de las preguntas).

Todos los ficheros necesarios para realizar este proceso y poder personalizar los juegos con los temas que cada una quiera pueden descargarse desde el enlace : http://www.ctim.es/SeriousGames/

**Acelerar el tiempo de juego deseado.**

Muy a menudo, sobre todo si queremos usar la aplicación en el colegio, el tiempo que se puede dedicar al juego es limitado e interesa acelerarlo de alguna manera para poder hacer las partidas más rápidas. Hemos abordado este problema agregando al principio del juego una opción (versión rápida) que al activarla se producen los siguientes cambios en el desarrollo del juego :

1. Los números de los dados se mueven al azar entre 4 y 9 en lugar del habitual 1 y 6.

2. Para llegar a la meta, no se requiere un número exacto al tirar los dados (como sucede normalmente en la Oca y el Parchís)

3. En el caso de parchís, cada equipo juega con dos fichas (en lugar de 4).

Otra opción que hemos añadido para acelerar el juego es que en lugar de tener que hacer clic para tirar el dado, se hace de forma automática cada vez que es necesario.

Con estas opciones se consigue acelerar considerablemente la duración de las partidas.

# RESULTADOS

El principal objetivo que teníamos en mente cuando diseñamos la aplicación fue tratar de cubrir el máximo número de usuarios potenciales. Esto es, la aplicación debe funcionar en cualquier sistema operativo, en cualquier arquitectura de hardware y con la opción de ser descargado para ser ejecutado sin conexión a Internet.

De acuerdo con estos requisitos, decidimos implementar la aplicación utilizando HTML (W3C, 1999), javaScript y jQuery (York, 2011). HTML y javaScript se pueden ejecutar en cualquier navegador web que proporcione un intérprete de javaScript. En la actualidad, la mayoría de los navegadores proporcionan esta funcionalidad. jQuery es una biblioteca de javaScript diseñada para simplificar el desarrollo de varios efectos dinámicos en una página HTML.

A continuación se muestran algunas imágenes ilustrativas de la interfaz de usuario que hemos diseñado para ejecutar la aplicación.

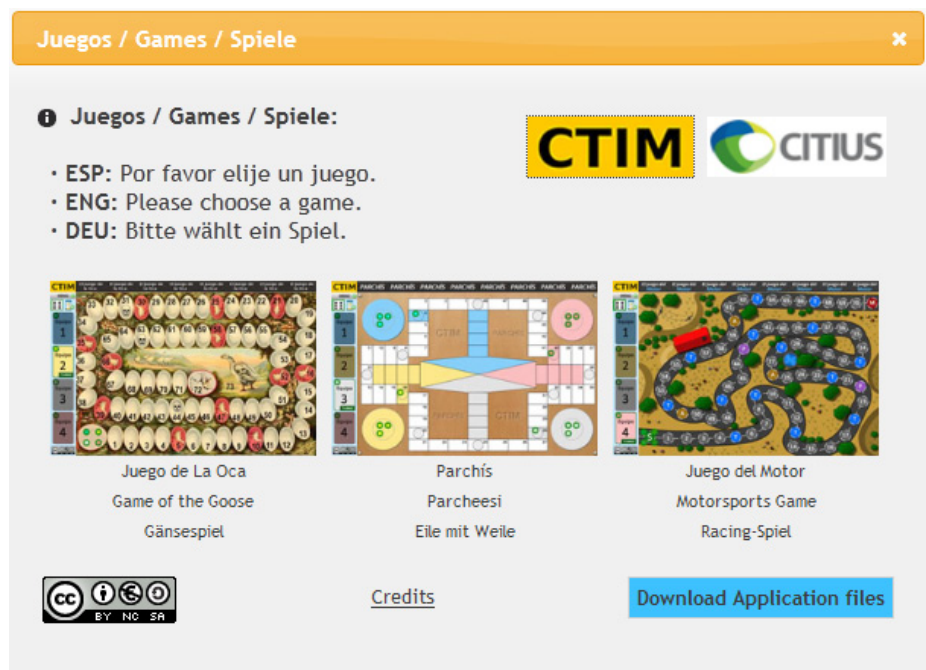

Ilustración 1: Cuando se inicia la aplicación se muestra un menú para
elegir el juego, el idioma, y si se desea descargar la aplicación.

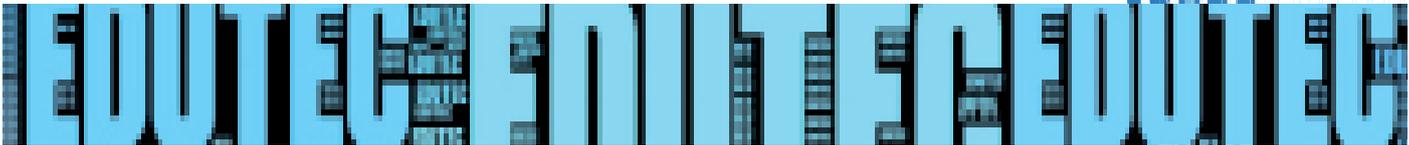

Ilustración 2: Al elegir el juego aparece una ventana para elegir los

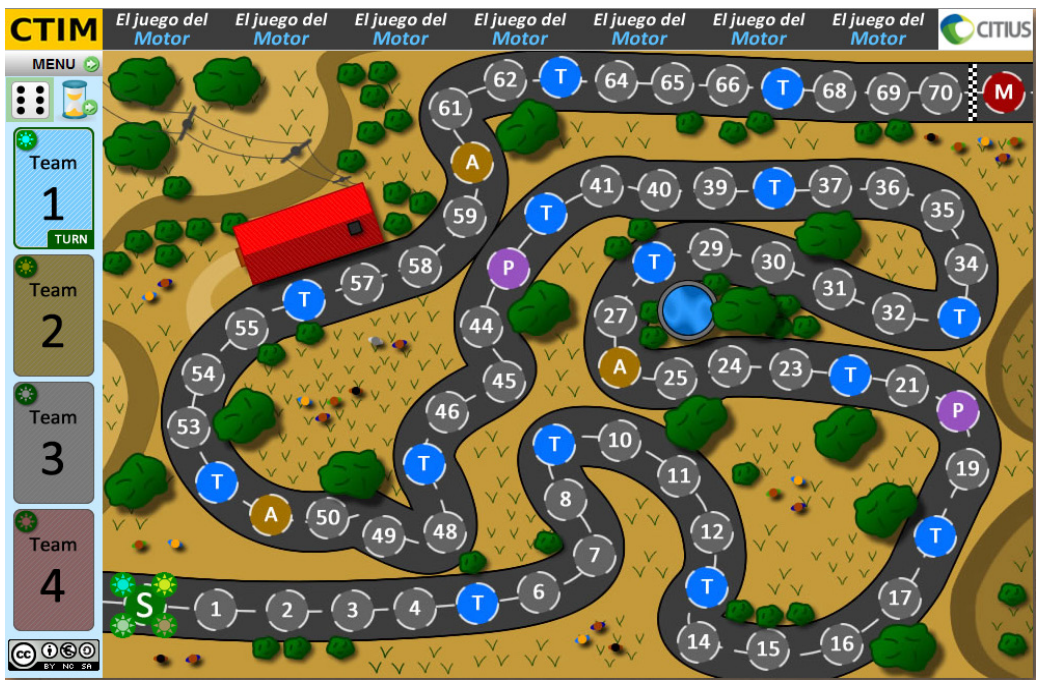

Ilustración 3: Se ilustra el Juego del Motor. A la izquierda aparece

el dado y la identificación de los equipos.

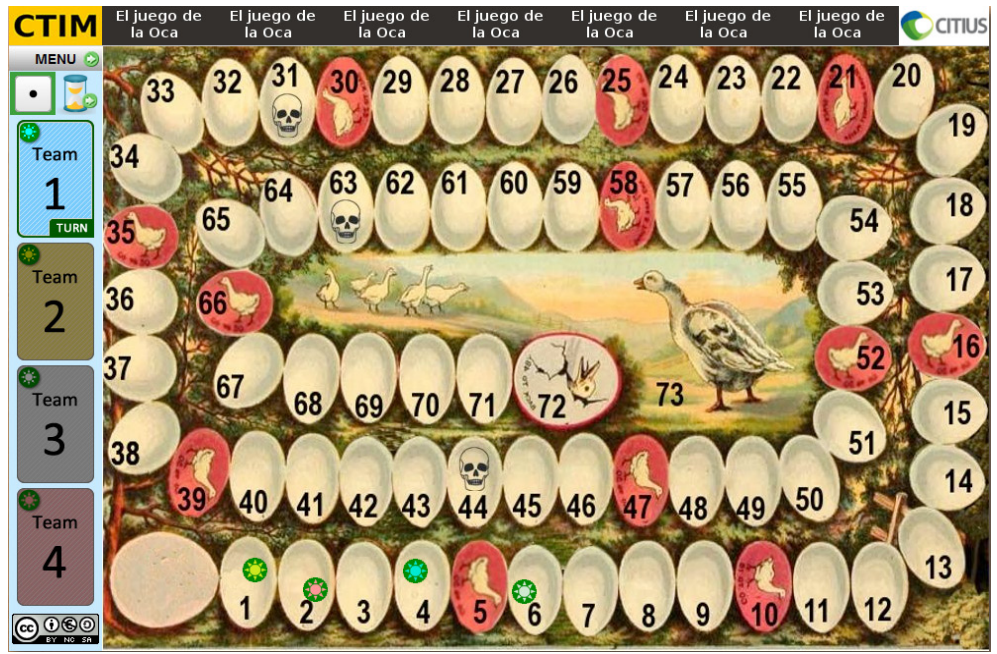

Ilustración 4: El Juego de la Oca

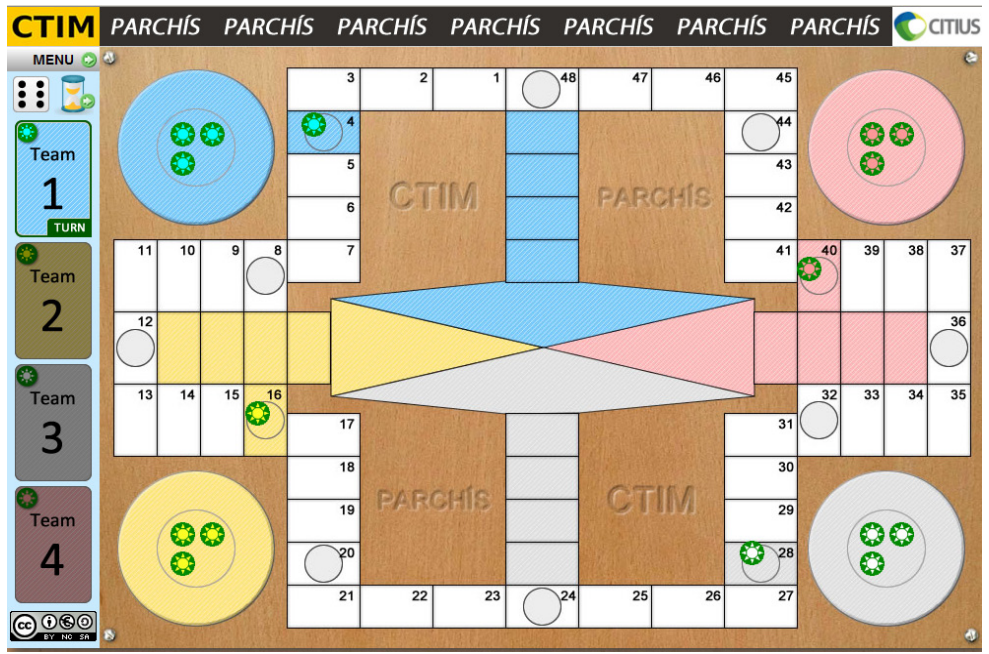

Ilustración 5 : El Juego del Parchís.

4 CONCLUSIONES

Hemos diseñado e implementado nuevos juegos educativos añadiendo funcionalidades educativas a los juegos de mesa clásicos.

Al principio de los juegos se eligen temas sobre los que se hacen preguntas a los jugadores/equipos a las que tienen que responder correctamente para que sus fichas se muevan. Las preguntas son escogidas al azar de la base de datos de preguntas.

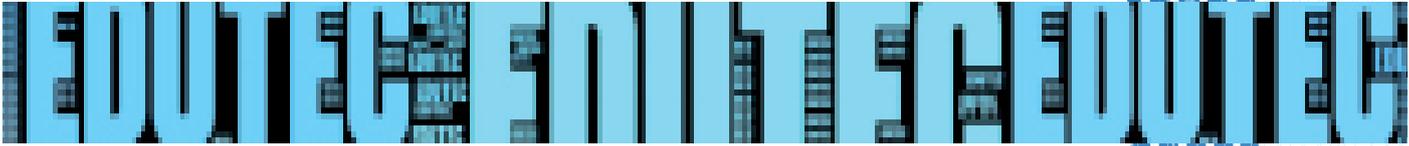

Hemos utilizado los juegos clásicos de la Oca y Parchís y diseñado uno nuevo: el Juego del Motor. Lo hemos implementado utilizando HTML, javaScript y jQuery para cubrir un número máximo de usuarios potenciales.

La aplicación software se distribuye bajo la licencia Creative Commons Reconocimiento-No comercial-Compartir bajo la misma licencia 3.0 Unported.

Los usuarios pueden añadir fácilmente nuevos temas y preguntas a la aplicación.

Creemos que este nuevo enfoque para añadir funcionalidades educativas a los clásicos juegos de mesa podría ser muy útil a nivel de la escuela para aprender muchos temas diferentes. Por otra parte, la posibilidad de añadir imágenes a las preguntas amplía el número de posibles temas que se pueden tratar.

# BIBLIOGRAFÍA